\documentclass[12pt,preprint]{aastex}

\shorttitle{Three stage model for GRB inner engine...}
\shortauthors{Staff et al.}

\begin{document}

   \title{A three stage model for the inner engine of Gamma Ray Burst: 
Prompt emission and early afterglow}

\author{Jan Staff and Rachid Ouyed} 

\affil{Department of Physics and Astronomy, University of Calgary,
2500 University Drive NW, Calgary, AB T2N 1N4, Canada}

\and 

\author{Manjari Bagchi \altaffilmark{1}}

\affil{Tata Institute of Fundamental
Research, Homi Bhaba Road, Colaba, Mumbai 400005, India}

\altaffiltext{1}{Department of Physics and Astronomy, University of Calgary,
Canada AB T2N 1N4}

\date{}

\begin{abstract}
We propose a new model within the ``Quark-nova'' scenario to 
interpret the recent observations of early afterglows of long Gamma-Ray
Bursts (GRB) with the Swift satellite. This is a three-stage model within
the context of a core-collapse supernova. Stage 1 is an accreting (proto-)
neutron star leading to a possible delay between the core collapse and the
GRB. Stage 2 is an accreting quark-star, generating the prompt GRB. Stage 3,
which occurs only if the quark-star collapses to form a black-hole, consists
of an accreting black-hole. The jet launched in this accretion process
interacts with the ejecta from stage 2, and could generate the flaring
activity frequently seen in X-ray afterglows. This model may be able to account
for both the energies and the timescales of GRBs, in addition to
 the newly discovered early X-ray afterglow features.
\end{abstract}


\section{Introduction}

\begin{figure}
\centering
\includegraphics{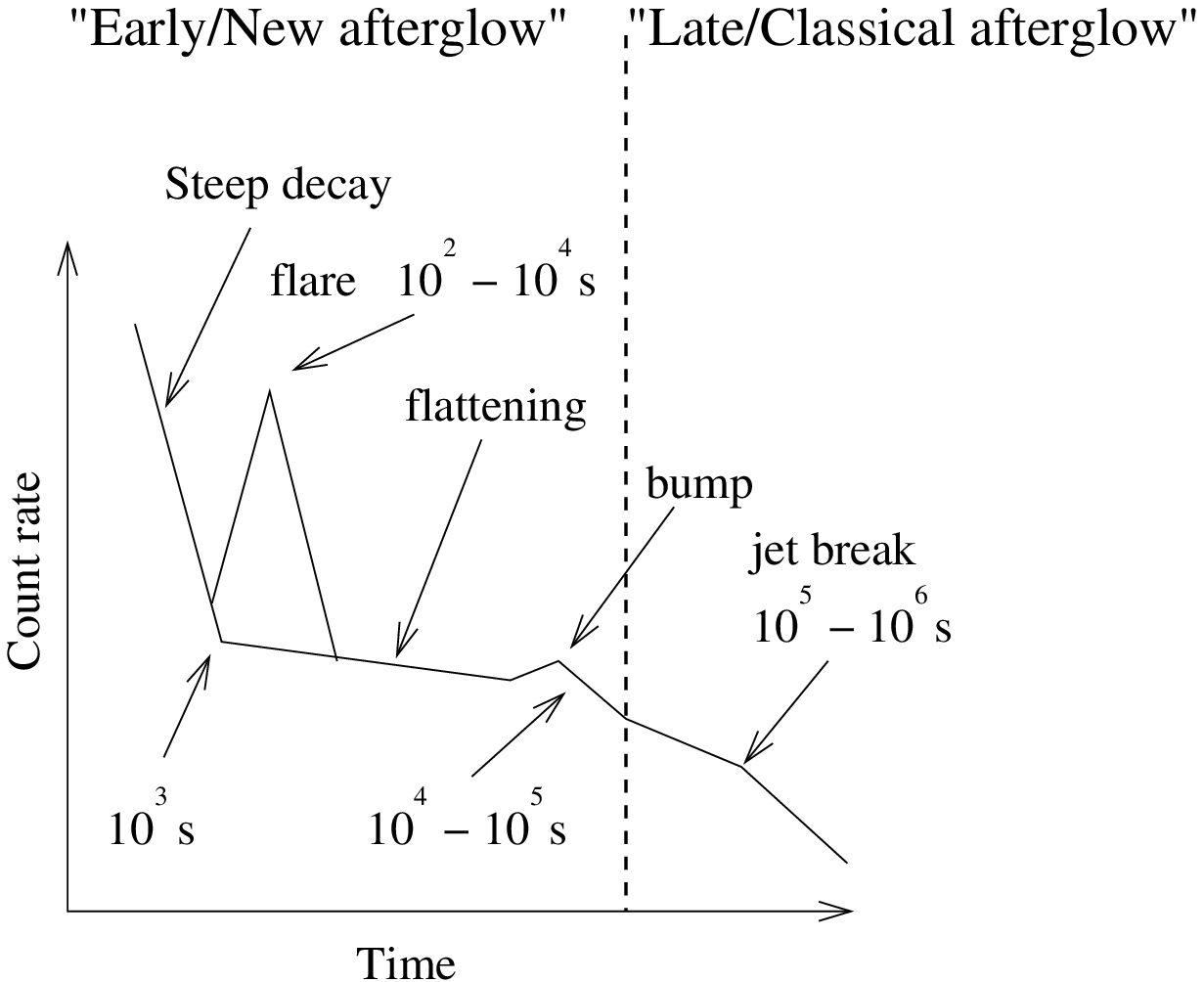}
\caption{Generic X-ray afterglow \citep[e.g.][]{zhang06}. At early
times ($t<1000$ seconds), a 
steep decay is often seen, followed by a flattening starting at
$t=10^3$ and lasting  for about $10^4-10^5$ seconds. Thereafter a 
steeper decay, and after about a day or so a new break is seen. 
The last break is termed ``jet break''.
On top of this one or more 
flares are often observed between $t=10^2-10^4$ seconds, and one or more
bumps can be seen between $t=10^4$ seconds and $t=10^5$ seconds. {\bf NOTE:}
Not all features are seen in every bursts. Before (the launch of the) Swift
satellite, the part of the figure including the steep decay, the flares and
the flattening was essentially unobserved.}
\label{genericfigure}
\end{figure}

With the launch of the Swift satellite \citep{gehrels04}, 
observations of early X-ray afterglows from gamma ray bursts (GRBs) 
have become possible (sometimes as early as 80 
seconds after the GRB trigger). This led to some 
surprising observations, most notably, the existence of one or more (sometimes
giant) flares \citep[first discussed by][]{burrows05} in the early X-ray afterglow (after a few hundred seconds to 
several thousand seconds) whose rapid variability has been interpreted as
the inner engine being active much longer than the 
duration of the GRB itself \citep[see for instance][]{zhang06}. 
Figure~\ref{genericfigure} shows a canonical
X-ray afterglow light curve. The early X-ray afterglow often has a very
steep power-law decay, lasting up to about a thousand seconds, followed by a 
flattening
of the light curve, which lasts for $10^4-10^5$ seconds. This is often
overlayed with flares and bumps in the light curve. Thereafter, 
a ``classical'' decaying afterglow is seen, as it was known from the 
pre-Swift era. It should be noted that not every bursts exhibits all of 
these features. The flare(s) and the flattening of the light curve are
likely due to extended inner engine activity, and, this extended inner engine
activity is what we focus on in this paper.

The inner engine for the GRB in our model is an 
accreting quark-star formed shortly (hours) after the core collapse in 
a massive star. This process operates
in three steps: i) the quark nova \citep{ouyed02}, where the core
is converted to quark matter resulting in  mass ejection, ii) fall back 
material from the supernova together with some of
the matter ejected during the quark nova can form an
accretion disk around the quark-star. Accretion onto the quark-star 
launches a jet that will overtake the shell
ejected by the quark nova and is a possible location for the
prompt X-ray  emission. Internal shocks within the jet produce the
GRB, iii) if enough matter is accreted onto the quark-star, it
will collapse and form a black-hole. Continued accretion onto the
black-hole can lead to an ultra-relativistic jet. Interactions
between this jet and the jet from the quark-star create shocks which 
lead to the flaring activity frequently seen in X-ray afterglows. 

In this paper we assume that the gamma-ray emission is produced by internal
shocks. The afterglow is produced when the merged shell creating the 
gamma-ray emission interacts with the external medium in an external shock.

It should be emphasized that quark stars are hypothetical objects, that
have not been observed. They have been discussed theoretically for more than
thirty years \citep[see for instance][]{itoh70}. Two requirements are needed
for quark stars to form. First the strange quark
matter (SQM) hypothesis must hold true. This hypothesis states that SQM has a
lower energy state than $^{56}Fe$, thus being the ground state of hadronic
matter. Quark stars are
usually thought to form from neutron stars whose central density increases
above a critical density. Only the more massive neutron stars are thought to
be able to reach this density \citep{staff06}. Second, 
the SQM EOS must support a rather massive quark star
(the mass will probably not change much in converting from a neutron star to
a quark star). Optionally, a quark star can form directly in the collapse of
the iron core in massive stars, in which case the second requirement might
be relaxed.

In \S~\ref{modelsection} we present our model for an inner engine for
GRBs in three stages, followed by a discussion of the energies and
timescales of the inner engine in \S~\ref{ietimeenergiessection}. Then
in \S~\ref{bhqsintsection} we discuss what happens when the jet from
the black-hole and the quark-star interacts followed by a discussion of the
formation of the different features in the X-ray afterglow in
\S~\ref{lcfeaturesection}. We devote \S~\ref{flaresbumpsection} to
correlations of features seen in the early afterglow light curve,
and then in \S~\ref{casestudysection} we apply our model to
different observed GRBs.  Then in \S~\ref{discussionsection} we discuss
our model and finally we summarize our model in
\S~\ref{summarysection}.

\section{The three stages of activity}
\label{modelsection}

\begin{figure}
\includegraphics[width=\textwidth]{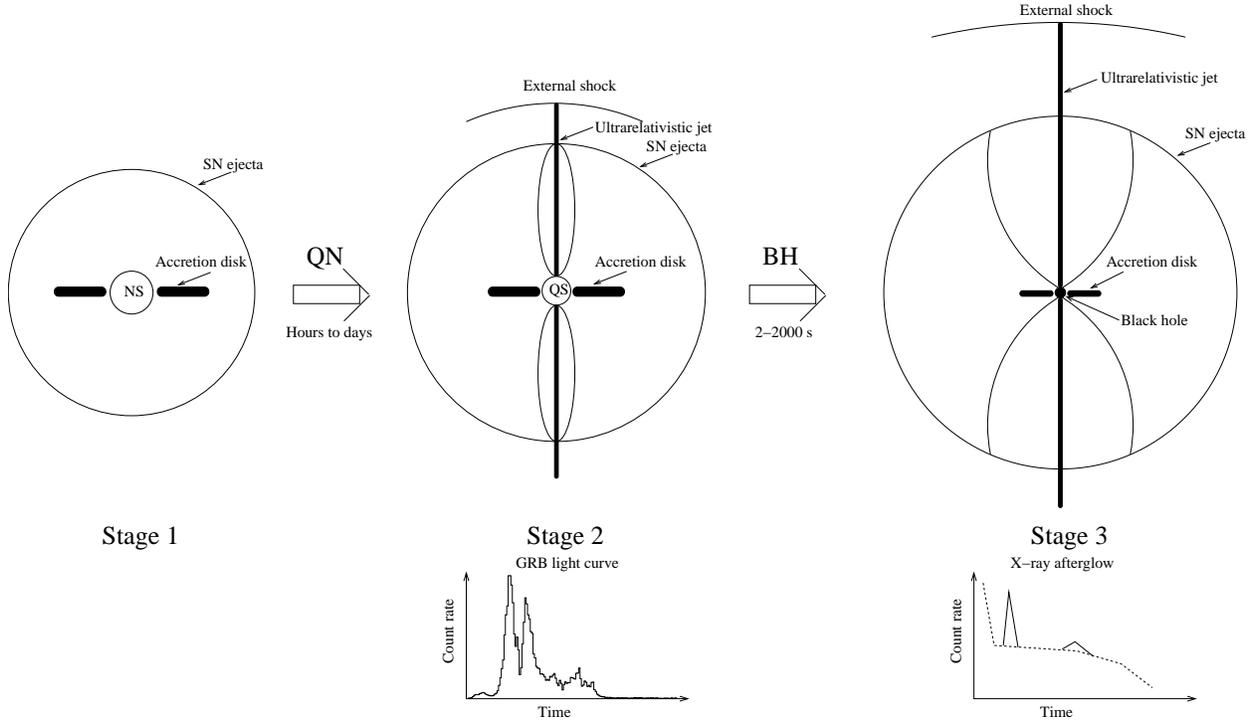}
\caption{The three stages in a GRB inner engine. {\it Stage 1:} A 
neutron star is formed in the core collapse of a massive star. The neutron star can later
collapse into a quark-star due to spin-down or accretion. If this takes
a long time, it will cause a delay between the supernova and the GRB.
{\it Stage 2:}
A quark-star in the CFL phase is formed, either directly 
in the supernova, or from the neutron star in Stage 1. Accretion onto this
star heats the star, leading to emission of photons as the main cooling
mechanism. If the star is heated above the plasma frequency of about 7.7 
MeV (ORV), the photons can escape the
star and will interact with the accreting matter. The infalling matter is
ejected, leading to a halt in the accretion process. When the star cools
down below 7.7 MeV, accretion can be resumed. This creates episodes of 
accretion and ejection. Internal shocks created from colliding shells in the
ejected material create the GRB. {\it Stage 3:} At later stages in the
accretion process the accretion may not be strong enough to heat the star
above 7.7 MeV. This prevents ejection of material, and hence terminates the 
GRB. However, the accretion will continue. If the star accretes too much
matter, it will collapse to a black-hole. This dramatically changes the 
accretion, and launches an ultrarelativistic jet. This jet will interact
with slower parts of the jet from the quark-star, leading to internal 
shocks. This creates the flares often seen in the early X-ray afterglow.
When the jet from the black-hole collides with the external shock, a bump
is observed on the light curve.}
\label{threestagesfigure}
\end{figure}

In this section, we discuss the three stages in our model for the inner
engine for long GRBs (see Fig.~\ref{threestagesfigure}). These stages involve a neutron star phase, a quark
star phase, and a black hole phase. The goal is to explain features (e.g. 
flares and flattening in early X-ray afterglow) observed in long GRBs as a
result of interactions between the neutron star jet, quark star jet, and the
black hole jet. In essence the interaction between these jets is more
generic and could be applicable to other multi-stage models involving other
engines.

\subsection{Stage 1: Neutron star as the inner engine}

The first stage in our model is the formation of a (proto-) neutron star
from the collapse of the core in the supernova. 
Several suggestions on how to generate GRBs from a newly formed neutron
star exist \citep[e.g.][]{usov92,kluzniak98,wheeler00,ruderman00,dai06}.
More elaborate models involve in addition a transition to a black hole.
Here we bring in an intermediate stage where the neutron star collapses to
a quark star (i.e. a quark nova) first. Our model is effectively a two-stage
model since we will be mostly focusing on the interaction between the
quark star ejecta and the black hole ejecta. We note 
however that the neutron star phase can lead to a short delay between the
collapse of the iron core and the GRB, which in our model occurs during
the last two stages which we describe next.

\subsection{Stage 2: Quark star as the inner engine}

 The neutron star will accrete fall-back material from the exploding star
 until its density reaches deconfinement density, whereby the star is
 converted to a strange quark-star. Alternatively, if the density of the
 compact object left behind in the core collapse is above deconfinement
 density, the quark-star can be formed directly in the collapse. The
 accreting quark-star will produce the GRB \citep[hereafter ORV]{ouyed05},
 which is discussed later. 

The lifetime of the neutron star will determine whether the GRB and 
supernova will be seen as simultaneous (or nearly
simultaneous) events, or as temporally separated events. The timescale
for the conversion of the neutron star can vary dramatically, depending on the accretion rate, the
spin-down rate, and how close the state of the neutron star is to a
deconfined state. Observations indicate that the core collapse and the GRB are
normally almost simultaneous events \citep{dellavalle06}, which means that
stage 1 in our model is usually short or not present at all.

The quark-star will be surrounded by an accretion disk, due to fallback 
material from the supernova, as well as quark nova material. As the star 
accretes matter, it will be heated up. The decay of Goldstone bosons 
producing photons is the main cooling mechanism in
this phase \citep{vogt04}. For temperatures above about 7.7 MeV 
these photons can escape the star, whereas for lower temperature they will
be absorbed by the star (see ORV). If the star heats up above 7.7 MeV, the emitted 
photons will interact with accreting material and eject it. This halts 
accretion until the star has cooled down below 7.7 MeV again. This 
way, episodes of accretion and ejection will occur.

The accreting material will follow the star's magnetic field lines towards the 
magnetic pole of the star. Hence, most of the ejected material will be 
ejected from the polar regions, and it will be collimated by the magnetic
field, i.e. it is a jet (ORV).
We assume an accretion rate of $\dot{m}\sim10^{-4}M_\odot~{\rm s^{-1}}$ onto the
quark star (similar to what is expected for neutron
stars, Fryer et al. 1996). 
Explaining the physical process that limits the accretion rate is beyond
the scope of this work and is left for future work. 
If the accretion rate is higher, we suggest a black hole is formed quickly.
This gives us a one stage model, as described in
\S~\ref{noqsstagesubsection}. 

We find that for typical values of the QS magnetic field, $B\sim10^{15}$ 
G\footnote{Recent work shows that $10^{15}$ G magnetic fields can be 
obtained during QS formation due to the response of quarks to 
the spontaneous magnetization of the gluons \citep[e.g.][and 
references therein]{iwazaki05}.},
the maximum accretion rate that can be channeled towards the polar
cap\footnote{The lack of observational evidence of a precessing jet
\citep{beloborodov00} is naturally explained in our model
since the magnetic field immediately after birth will align with the
rotational axis of the star \citep{ouyed06}.} to be
$\dot{m}\approx10^{-3} M_\odot~{\rm s^{-1}}$ (for higher accretion rates the
Alfv{\'e}n radius lies inside the star). For $B\sim10^{14}$ G, the accretion
rate must be smaller than $\dot{m}\approx10^{-5} M_\odot~{\rm s^{-1}}$ for the
accretion to be channeled to the polar cap by the magnetic field, too low for
the more energetic burst but suitable for lower energy bursts and very long
duration bursts.

\begin{figure}
\includegraphics[width=\textwidth]{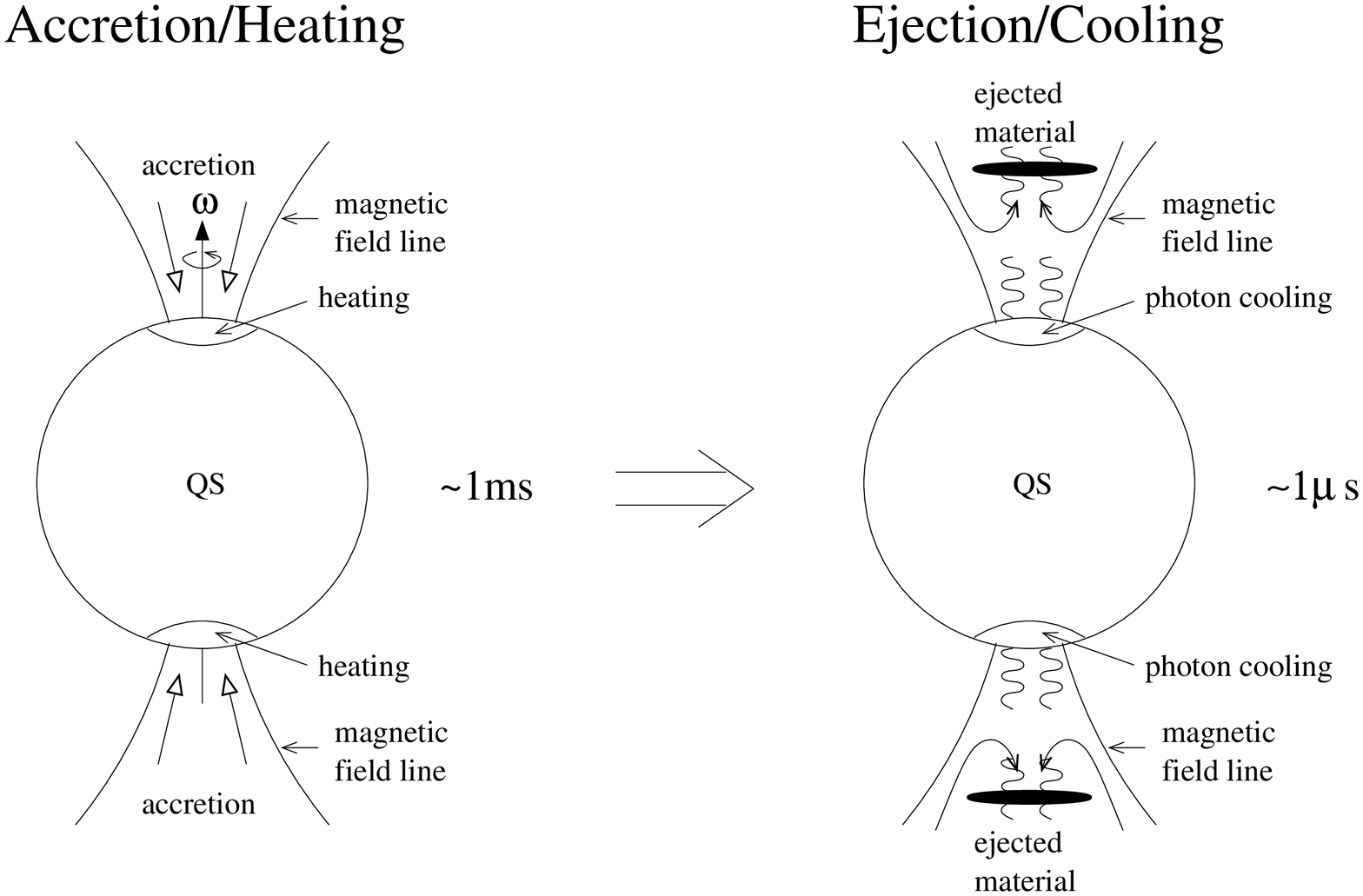}
\caption {Cartoon illustrating the jet launching mechanism in our model.
Infalling material follows magnetic field lines to the polar cap (magnetic pole
is at the same location as the geographic pole in a CFL star, see text), where
it heats the star. The timescale for heating the star is of the order 
milliseonds. The star then cools by emitting photons on a timescale of the
order of microseconds. These photons
interact with the infalling material, ejecting some of that with large Lorentz
factors.}
\label{acc-ej}
\end{figure}

Most photons will be emitted from the polar regions (where accretion heats 
up the star), and they will then interact with some of the accreting material
(see Fig~\ref{acc-ej}).
The Lorentz factor of the ejected matter is $\Gamma=\eta_{\rm QS}m_{\rm accr}
/m_{\rm ejec}$, with $\eta_{\rm QS}\sim0.1$ \citep{frank92}, $m_{\rm accr}$
and $m_{\rm ejec}$ are the accreted and ejected mass respectively (ORV). In
order to achieve Lorentz factors of the order 100, only a small fraction
($10^{-3}$) of the accreting material can therefore be ejected. Such a small
fraction can be realized since the cooling time (and therefore the period in
which photons are emitted) is of the order microseconds, whereas the heating
(accretion) time is of the order milliseconds (ORV). If we assume a steady
accretion, and that the photons eject an amount of mass
$m_{\rm ejec}\sim\dot{m}\Delta t_{\rm em}$ for each episode we find:
\begin{equation}
\frac{\Delta t_{\rm accr}}{\Delta t_{\rm em}}\sim\frac{m_{\rm eject}}{m_{\rm accr}}\sim10^{-3},
\end{equation}
$\Delta t_{\rm accr}$ and $\Delta t_{\rm em}$ being the time interval for accretion and 
ejection during one episode respectively.
However, it is likely that only a fraction of this material is ejected,
leading to higher Lorentz factors. This fraction might vary from episode to
episode, giving rise to varying Lorentz factors in the outflow. Internal
shocks created by colliding shells in the jet accelerate electrons that emit
synchrotron radiation observed as gamma rays, as in \citet{narayan92}.

In order to produce merged shells with an internal energy of about $10^{50}$
erg (which is generally required to explain the energies involved in 
GRBs) and Lorentz factors of more than a hundred \citep[to overcome the 
compactness problem; see e.g.][]{piran99}, the shells must be about 
$10^{-6}M_\odot$ or $10^{27}$ g each. 
If the Lorentz factors are too large (a few thousands), internal shocks
will occur too late and external shocks occur before the internal shocks
can take place.
This sets the upper and lower limits on the Lorentz factors required for 
GRBs. For $\eta_{\rm QS}=0.1$, this means that 
$m_{\rm accr}=10^{30}-10^{31}{\rm g}\sim10^{-3}-10^{-2}M_\odot$. A more
likely scenario is that in each episode less mass is ejected, and
several of these ejecta quickly merge to produce shells with mass
$\sim10^{27}$ g relaxing the constraints on $m_{\rm accr}$ to lower
values. These shell mergers will not be observed, as they occur
while the jet is still inside the exploding star.
 The most efficient conversion of kinetic energy to internal energy is
achieved when the masses of the shells are similar, and the ratio 
in Lorentz factor of the colliding shells are big, at least a factor two.

The duration of the accretion process depends on the mass of the disk and
the maximum temperature that the star can be heated to. ORV found that 
this process can last hundreds of seconds, and possibly thousands. We will
assume that the accretion eventually settles into a steady state. If this
accretion is not high enough to heat the star much above 7.7 MeV, the 
ejection will be halted or very limited and the GRB comes to an end. 
However, the inner engine is still actively accreting, 
until the disk has been depleted; this is unless the star turns into 
a BH before  disk depletion (see next subsection for more discussion).

ORV found that $10\%$ of the rest mass energy of the 
accretion disk being accreted can be used to power a jet. A quark-star can 
probably accrete up to $0.1M_\odot$ without collapsing to a black-hole.
Hence $0.1M_\odot c^2\approx 2\times10^{53}$ erg is the maximum
jet energy powered by accretion onto a quark-star. We assume a radiative
efficiency of $10\%$ in shell collisions during the QS phase. This
means that in our model only about $1\%$ (or about $2\times10^{51}$ erg) 
of the rest mass energy of the accretion disk surrounding the quark-star 
can be released as gamma rays in a GRB. Using a collimation angle of a few 
degrees (ORV), this corresponds to 
$E_{\gamma, {\rm iso}}\sim10^{54}$ erg.

The ejected shells will collide with each other as explained, and be
decelerated by the interstellar medium forming an external shock.
Acceleration of electrons in this
shock creates the afterglow (as in the internal--external shocks model). If
the outflow creating the afterglow is beamed with jet angle 
$\theta_{\rm jet}$, a jet break will be observed
as $\Gamma^{-1}>\theta_{\rm jet}$ \citep{rhoads99}. This usually happens after
about a day.

\subsection{Stage 3: Black-hole as the inner engine}
 
For a given temperature and equation of state (EOS), a mass-radius curve for
quark-stars has a maximum, i.e. there is a maximum mass for quark-stars.
As is the case for the maximum mass of neutron stars, this maximum mass
depends on the EOS which is still being studied. \citet{harko02} shows that
quark stars whose EOS can be approximated by a linear function of the
density has a Chandrasekhar limit based on degenerate quarks.
This point denotes an instability, and so if there is
accretion onto a star having this maximum mass, it collapses into a black
hole. 
We note however, that thermally induced 
instabilities can drive the collapse to a BH before reaching the Chandrasekhar
limit
\citep[for details see][]{bagchi06}.

When the quark-star collapses into a black-hole, the accretion process
changes dramatically. 
The accretion rate is of the order
$0.01$ to $10$ $M_\odot~{\rm s^{-1}}$ in a hyperaccretion disk around black
holes \citep{popham99}, much higher than around quark stars where surface
radiation and magnetospheric effects should in principle reduce the accretion rate
compared to the accretion rate onto black holes.
An ultrarelativistic jet is launched from the accretion process onto the
black hole as in \citet{devilliers05}. Interaction between this jet and the
slower
parts of the jet from the quark-star can generate flares seen in the early
X-ray afterglow and when this jet collides with the external shock, late
time bumps will be seen.
Internal shocks within the black hole jet itself can also occur, 
which likely would lead to flares in either X-ray or gamma-ray wavelengths. 
Since the duration of this jet is likely short, the width of these flares 
would also be short. This may add to the complexity of the observed light 
curve.

The duration of the accretion/ejection process depends on the mass of the
accretion disk after the quark-star to black-hole conversion 
and the accretion rate.  The disk is unlikely to be more than a few solar
masses, giving a maximum accretion time of about a hundred seconds. However
as found by
\citet{devilliers05}, 
accretion onto a black-hole is a much faster process (about 
$1M_\odot~{\rm s^{-1}}$), which 
limits the duration of the accretion process to a few seconds. This we
take as a typical timescale for the accretion rate onto black-holes for the
rest of this paper.

\section{Inner engines timescales and energies}
\label{ietimeenergiessection}
We can estimate the ratio between the duration of the inner engine in the
black-hole stage and in the quark-star stage as:
\begin{equation}
\frac{t_{\rm BH}}{t_{\rm QS}}=\frac{\frac{m_{\rm disk, BH}}{\dot{m}_{\rm BH}}}{\frac{m_{\rm disk, QS}}{\dot{m}_{\rm QS}}}
=\bigg(\frac{\dot{m}_{\rm QS}}{\dot{m}_{\rm BH}}\bigg)\frac{m_{\rm disk}-m_{\rm disk,QS}}{m_{\rm disk,QS}}
=\zeta_{\dot{\rm m}} \bigg(\frac{m_{\rm disk}}{m_{\rm disk,QS}}-1\bigg)
\label{timefraqeq}
\end{equation}
where $m_{\rm disk}=m_{\rm disk,QS}+m_{\rm disk,BH}$ and $\zeta_{\dot{\rm m}}$ parameterizes 
the ratio between the accretion rate onto a quark-star and a black-hole. 
As typical values we use the same accretion rate onto quark stars as for neutron stars, 
$\dot{m}_{\rm QS}\approx10^{-4} M_\odot~{\rm s^{-1}}$ \citep[in fact the accretion 
rate depends on the maximum temperature the star is heated to as in ORV]
{fryer96}
and $\dot{m}_{\rm BH}\approx1 M_\odot~{\rm s^{-1}}$ \citep{devilliers05}
giving $\zeta_{\dot{\rm m}}\approx10^{-4}$, 
and find that the duration of the BH era is much shorter 
than the QS era. 

The ratio between the energy produced by the inner engine in the
two stages can be found by:
\begin{equation}
\frac{E_{\rm BH}}{E_{\rm QS}}=\frac{\eta_{\rm BH} m_{\rm disk,BH}}{\eta_{\rm QS} m_{\rm disk,QS}}
=\zeta_{\rm m}\frac{m_{\rm disk}-m_{\rm disk,QS}}{m_{\rm disk,QS}}
=\zeta_{\rm m}\bigg(\frac{m_{\rm disk}}{m_{\rm disk,QS}}-1\bigg).
\label{energyfraqeq}
\end{equation}
For typical values we use $\eta_{\rm BH}\approx10^{-3}$ 
\citep[$m_{\rm ejec}/m_{\rm acc}=10^{-5}$ was found in][assuming  
$\Gamma=100$ this gives 
$\eta_{\rm BH}=\Gamma m_{\rm ejec}/m_{\rm acc}=10^{-3}$]{devilliers05}, 
and $\eta_{\rm QS}=10^{-1}$ \citep[similar to the efficiency in neutron 
stars,][]{frank92}. 
This gives $\zeta_{\rm m}=10^{-2}$.
Hence, most of the energy will
be output during the QS era, unless almost all of the disk ($>99\%$) is
accreted during the black-hole era.

We can now combine the two above expressions, to get a simple ratio:
\begin{equation}
\frac{t_{\rm BH}}{t_{\rm QS}}=\frac{\zeta_{\dot{\rm m}}}{\zeta_{\rm m}}
\frac{E_{\rm BH}}{E_{\rm QS}}.
\label{combinedeq}
\end{equation}

From Eq.~\ref{energyfraqeq} we find that if $10\%$ of a disk is accreted
onto a quark-star (which then collapses into a black-hole) and the rest is 
accreted into the black-hole, the energy output
in a jet from the quark-star is ten times larger than the energy output
in a jet from the black-hole.
In this case we see from Eq.~\ref{combinedeq} that the duration of 
the accretion process onto the quark-star is a thousand times longer 
than the duration of the accretion into the black-hole. 

The above discussion is valid for the inner engine. We will now try to
relate that to observations.
We assume a direct relation between the energy output from the inner
engine in the black-hole and quark-star phases and the observed energies 
from the two phases:
\begin{eqnarray}
\nonumber E_{\rm BH, obs}&\simeq\langle \zeta_{\rm sh,BH}\rangle E_{\rm BH}\\
E_{\rm QS, obs}&\simeq\langle \zeta_{\rm sh,QS}\rangle E_{\rm QS},
\end{eqnarray}
where $\langle\zeta_{\rm sh,BH}\rangle$ and $\langle\zeta_{\rm sh,QS}\rangle$ 
is the energy conversion 
efficiency in the shocks averaged over the entire duration of the shock 
activity \citep[e.g.][]{kobayashi97}. For multiple shocks, we 
assume this 
efficiency to be the same for the jet from the black-hole and the 
quark-star, hence $\langle\zeta_{\rm BH}\rangle\simeq\langle\zeta_{\rm QS}\rangle$.
Using this, we obtain the following ratios between the observed energies
and the accreted masses:
\begin{equation}
\frac{E_{\rm BH, obs}}{E_{\rm QS, obs}}
\simeq\frac{E_{\rm BH}}{E_{\rm QS}}
\simeq\zeta_{\rm m}\bigg(\frac{m_{\rm disk}}{m_{\rm disk, QS}}-1\bigg)
\end{equation}
To a first approximation we can assume that $E_{\rm QS, obs}$ is the 
observed GRB energy, and $E_{\rm BH, obs}$ is the observed energy in 
flares and bumps in the X-ray afterglow.

The observed energies in X-ray flares and bumps compared to the observed 
energies released in gamma rays varies a lot in different
bursts (from no flares or bumps to flares with fluence equal to the 
GRB fluence). If we take
\begin{equation}
\frac{E_{\rm BH, obs}}{E_{\rm QS, obs}}\simeq 10^{-2}
\end{equation}
then, in our model, it would imply that
\begin{equation}
m_{\rm disk, BH}\simeq m_{\rm disk, QS}.
\end{equation}
In other words, an equal amount of mass is accreted onto the quark-star
and the black-hole.
This holds true if we neglect the energy carried by the slow shells from 
the quark-star, causing the flattening of the early X-ray afterglow
light curve. This is a reasonable assumption, since the energy in the 
flattening is less than in the GRB itself.

As for a direct comparison of the time scale of the inner engine to the
observed time scale, this is not possible since this is rather dictated by 
the complex interaction between the black hole ejecta and the quark star 
ejecta as we show next.

\section{Interaction between black-hole ejecta and quark-star ejecta}
\label{bhqsintsection}

\begin{figure}
\includegraphics{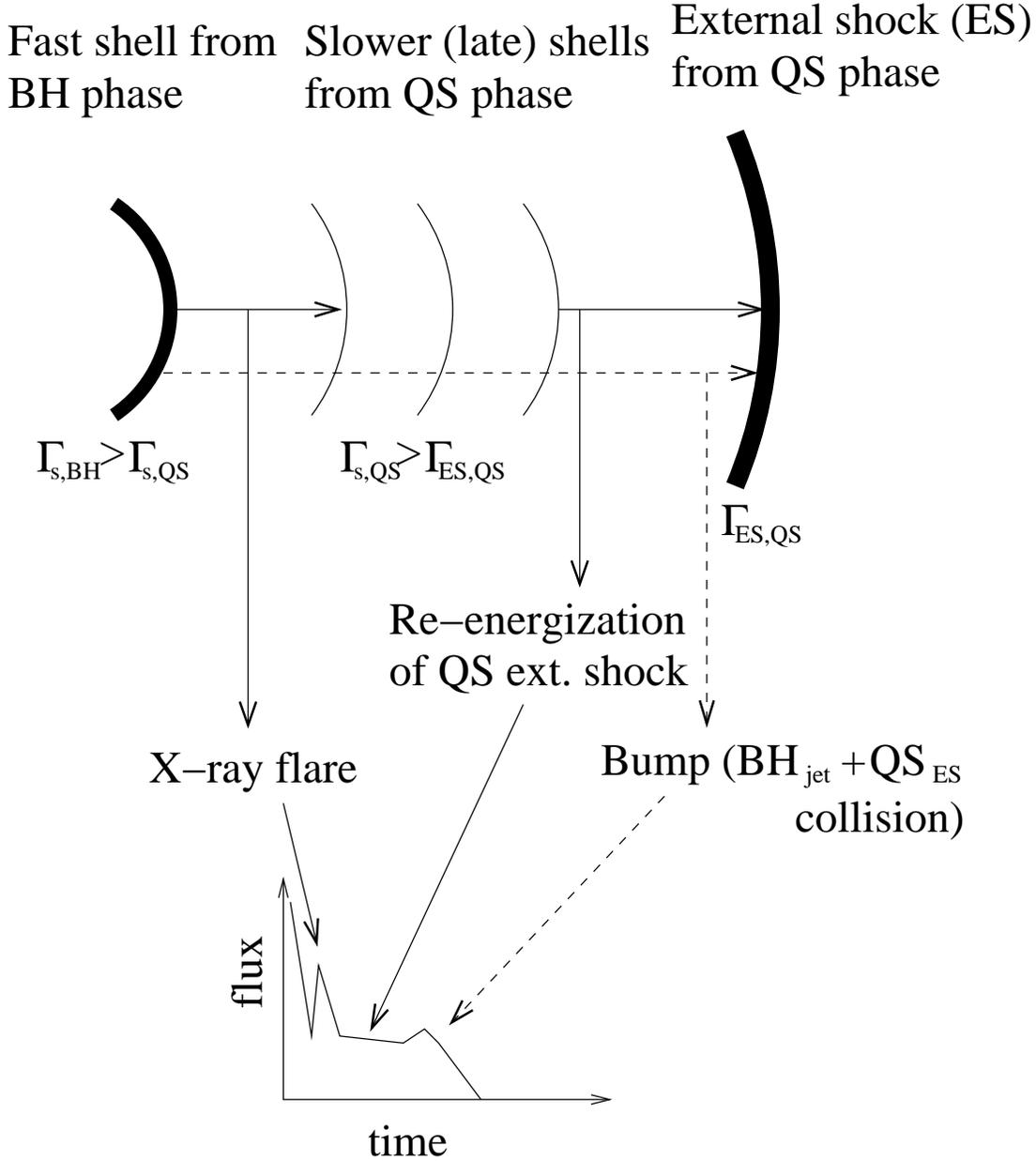}
\caption{Illustration of the mechanism leading to X-ray flares and flattening 
of the X-ray afterglow light curve. In both cases, the quark-star needs to
emit low-Lorentz factor shells in the late stages of shell emission. 
When these shells collide with the external shock from the GRB, that shock
is re-energized and a flattening of the light curve results. If the quark
star accretes too much mass, it will collapse to a black-hole. The jet
emitted in the
accretion process onto the black-hole can be massive and have a
high Lorentz factor. When this outflow collides with a slow shell from
the quark-star, an X-ray flare results. If this merged shell collides
with more low-Lorentz factor shells from the quark-star, more flares will
result. A bump can be seen when the black-hole ejecta collide with the 
external shock.}
\label{flare-flattening}
\end{figure}

Figure~\ref{flare-flattening} illustrates the process of flattening the
X-ray light curve and generating X-ray flares and bumps. If the quark-star
emits low-Lorentz factor shells at late stages of shell emission, the shell
from the black-hole can catch up with the last shell and produce an X-ray
flare via internal shocks. This merged shell may be capable of colliding
with other slow shells from the quark-star, whereby more X-ray flares will
be seen, an idea bearing resemblance to what is discussed in
\citet{zou06}. The energy output from the collisions depends on the
difference in Lorentz factor of the colliding shells. This is likely highest
in the first collision. However, the energy also depends on the masses of
the colliding shells, so the energy output is not necessarily highest in the
first X-ray flare (see eqs.~\ref{einteq} and
\ref{gammameq}).

If the mass and Lorentz factor of the merged shell producing the
X-ray flare(s) are high enough, a bump can be seen in the X-ray 
afterglow light curve as this shell collides with the external shock
from the GRB. 

The pulse width of the X-ray flares are generally longer than the width
of the flares in the GRB. There are also indications that later occurring 
flares are wider than flares occurring early \citep{zhang06}. Qualitatively
this can be understood from Eq.~\ref{pulsewidth} in the Appendix. 
The width of the burst
is proportional to the separation of the colliding shells. The shells 
creating the GRB were all produced by the quark-star in a fairly short
time, whereas the black-hole shell was created long thereafter. 

Our model seems capable of explaining why later flares have a longer
duration. From Eq.~\ref{pulsewidth} it is seen that the duration (for equal
mass shells)
 depends on the initial
distance between two shells. Since the flares in our model are all due to
the jet from the black-hole colliding with slower parts of the quark-star
jet, later flares are created by shells that were farther away from the
black-hole jet initially. This leads to longer duration.

The flattening of the X-ray light curve is due to lower Lorentz-factor
shells emitted in the later stages of the quark-star jet. These shells
will catch up with the external shock from the GRB at later stages, slowly
re-energizing the external shock. This is reminiscent of the refreshed
shocks scenario in \citet{sarimeszaros00}. Hence the re-energization is 
independent of the black-hole. The flare does depend on the black-hole
formation, and also on the fact that the quark-star jet emitted slower
shells. A flare does not necessarily lead to flattening, as only one slow
shell from the quark-star is needed to produce a flare, whereas a more
continuous sequence of low-Lorentz factor shells is needed to produce the
flattening.

The external shock will be decelerated by the surrounding medium. A low
Lorentz factor shell will catch up with the external shock when this has
decelerated to a comparable Lorentz factor. The range in Lorentz factors of
the slow part of the quark-star jet is necessary to explain a given flattening
therefore depends on the deceleration of the external shock, which is not
well known. Following \citet{falcone06}, a slow shell with Lorentz factor of
about 20 will catch up with the external shock after $t=10^4$ seconds,
assuming a uniform external medium with density $n=10~ {\rm cm}^{-3}$. A
shell with a Lorentz factor $\Gamma\sim9$ will catch up with the external
shock after $t=10^5$ seconds.

\section{Light curve features: {\it timescales and energies}}
\label{lcfeaturesection}

\begin{figure}
\includegraphics[width=0.19\textwidth]{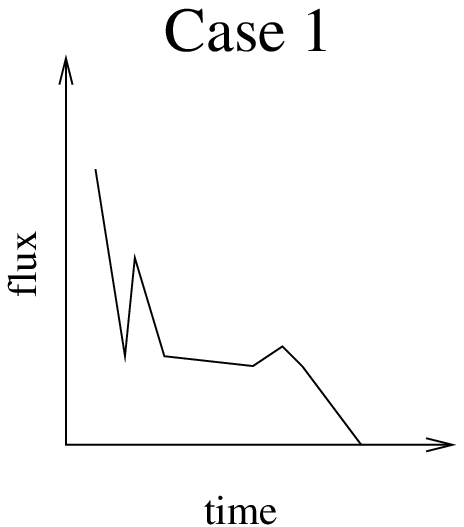}
\includegraphics[width=0.19\textwidth]{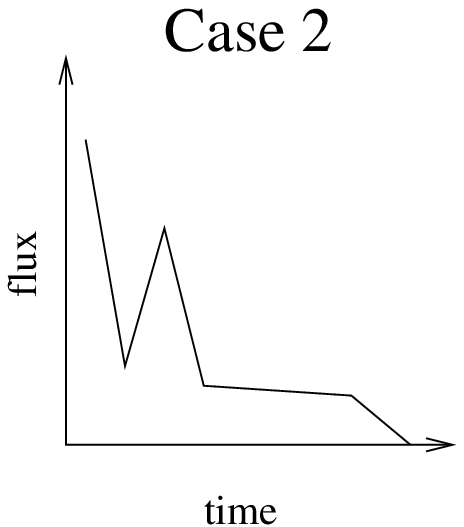}
\includegraphics[width=0.19\textwidth]{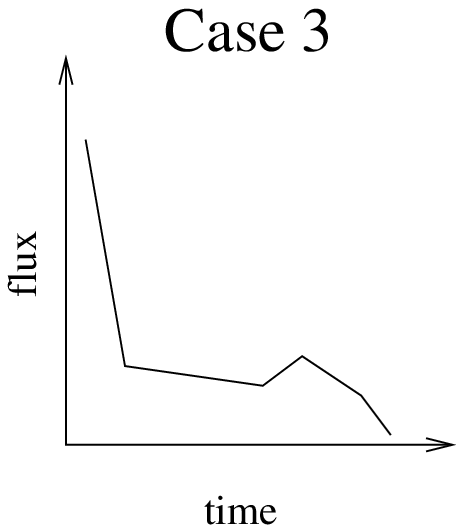}
\includegraphics[width=0.19\textwidth]{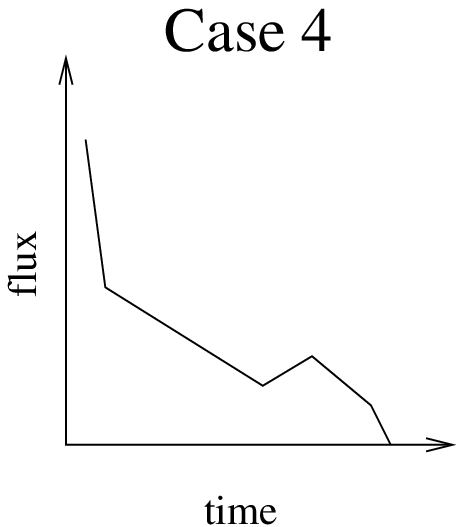}
\includegraphics[width=0.19\textwidth]{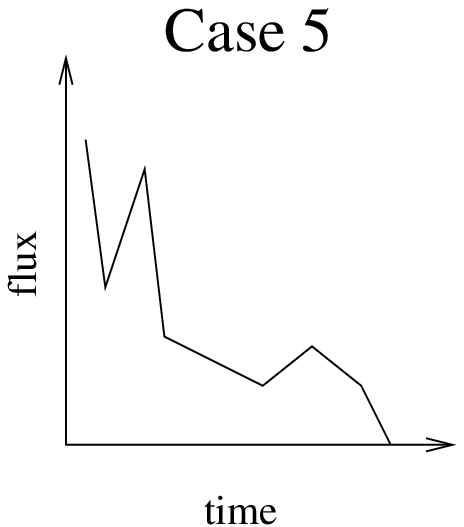}
\includegraphics[width=0.19\textwidth]{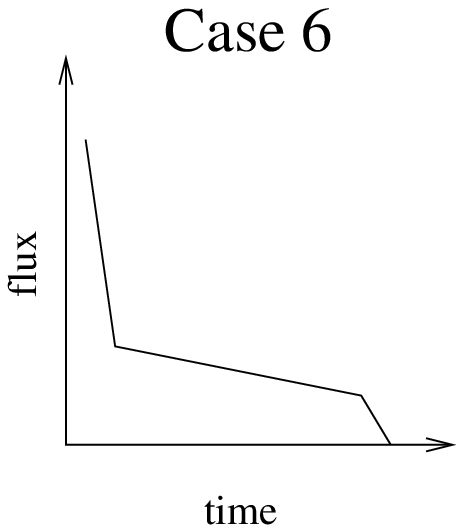}
\includegraphics[width=0.19\textwidth]{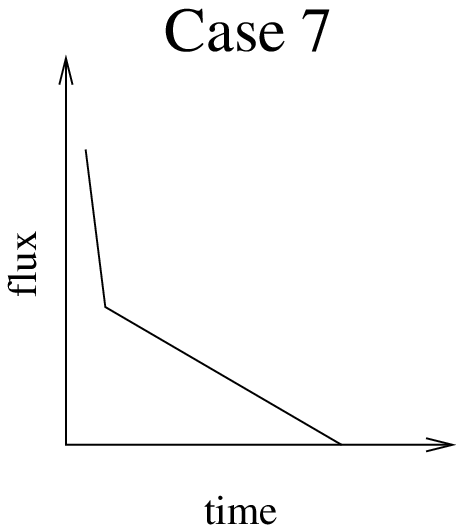}
\includegraphics[width=0.19\textwidth]{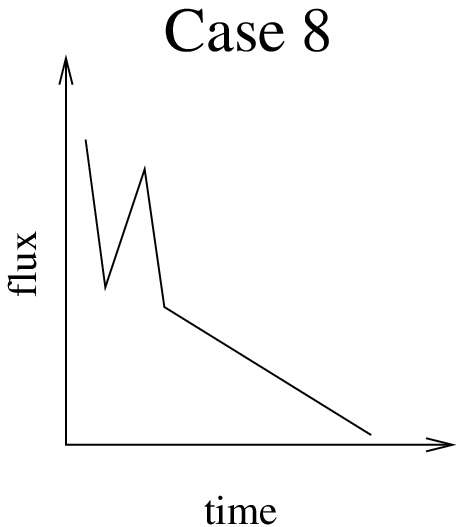}
\includegraphics[width=0.19\textwidth]{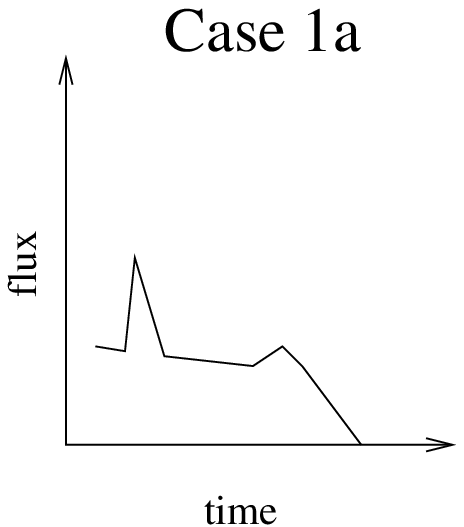}
\caption{The differeant kinds of light curves derived from our model.
Case 1 is also illustrated in Fig.~\ref{flare-flattening}. There are
flares, flattening and a bump. This occurs when the quark-star jet
produces slow shells that can re-energize the external shock. The black
hole jet interacts with some of these shells and creates flares. When the
merged black-hole jet collides with the external shock, the bump is formed.
{\it Case 2}: We 
see one or more flares and a flattening of the light curve, but no bump. 
This is because the shell from the black-hole is slowed down so much 
by the other shells from the quark-star that it reaches the external shock at 
very late times. {\it Case 3}: Flattening and a bump, but no flare. The shell
from the black-hole did not collide with any shell and therefore did not
create a flare. It creates a bump when it collides with the external shock.
{\it Case 4}: No flare, no flattening, but a bump. The quark-star jet
did not contain any late time shells, so there were no flattening and no
flare created. As in Case 3, the black-hole jet creates a bump when it 
collides with the external shock. {\it Case 5}: Flare and bump, but 
no flattening. There
are only a few late time shells emitted from the quark-star. The black
hole jet collides with these and creates flares. When this merged shell
collides with the external shock a bump is seen. No late time shells
can flatten the light curve. {\it Case 6}: No flares, no bump but flattening.
The quark-star emits late time shells that flattens the light curve. No
black-hole was formed. {\it Case 7}: No flares, no bump, no flattening.
No black-hole was formed, and the quark-star did not emit any late time
shells. {\it Case 8}: Flares, but no bump and no flattening. This
indicates that a black-hole which launched an ultrarelativistic jet 
was formed, but in the process of forming the flares this jet was
slowed down enough so that it cannot energize the external shock to generate
a bump. {\it Case 1a:} Same as case 1, but no early steep decay. {\it All}
cases from 1 to 8 can exist without the early steep decay as illustrated
in case 1a where the decay is absent for case 1.
{\it Note that a possible break in the light curve due to collimation of 
the outflow producing the GRB and the afterglow will occur at later 
times than what is shown in this figure.}}
\label{sixcases}
\end{figure}

In this section we first summarize the light curve features that our
model can produce (see Fig.~\ref{flare-flattening}), and then we
discuss the different possible lightcurves (Fig.~\ref{sixcases}).

{\it Steep decay:} The steep decay commonly seen in the early afterglow
($t\approx 10^2-10^3$ seconds) is due to the curvature effect 
\citep{kumarpanaitescu00}. This is prompt
gamma radiation from parts of the jet directed at an angle relative to the
line of sight. Because of the large Lorentz factors involved, most of
the radiation is beamed with a beaming angle $\Gamma^{-1}$. Some of it
will be directed to angles outside of $\Gamma^{-1}$, but then softened
by a Doppler factor. This is why what is seen as gamma rays along the
line of sight is seen as X-rays at an angle from the line of sight. The
steep decay continues until the luminosity from the external shock is
dominant. This effect is dominant only because the central engine does
not produce any visible activity, and the external shock is too weak to
be seen.

{\it Flattening:} The flattening of the light curve is due to the external
shock being re-energized by low Lorentz factor shells emitted from the
quark-star. These shells catch up with the decellerating external shock. These
shells were among the last ejected from the quark star.
It is important to note that there is not a one-to-one ratio for
the duration of the inner engine and the observed radiation in this case.
The inner-engine ejection of shells could last for some hundred seconds,
whereas the re-energization can last for $10^4-10^5$ seconds.

{\it Flares:} The flares are caused by interaction between a jet launched
by a black-hole and the slower parts of the ejecta from the quark-star.
In order for flares to occur, the quark-star must have accreted enough
matter that it collapses to a black-hole. 
The shell ejected by the black hole can be very massive, $m_{\rm
BH}\sim10^{28}$ g and have a very large Lorentz factor. The shells ejected
in the quark-star phase are likely less massive, however some of these might
merge into massive ones before the black-hole shell interacts with them.
In this case more than $10^{51}$ erg can be converted in one
flare. The pulse width depends on the initial separation of the shells and
on the width of the rapid shell \citep{kobayashi97}. The black hole shell can
be fairly wide initially, and it is not unlikely that it spreads more before
it collides with the quark-star shell. Finally, we mention recent work by 
\citep{liang06} where a zero time point (right before the beginning of the
flare) was taken as a signature of the reactivation of the engine. If true,
this observation suggests that the engine reactivates more than once which 
cannot be reconciled with our model. We will tackle this issue elsewhere.

{\it Bump:} The bumps seen at late times \citep[$t\sim10^4-10^5$ seconds as in
GRB 050502B;][]{falcone06} are the result of the black-hole jet or the
merged jet colliding with the external shock. The black-hole jet may or may
not have interacted with shells from the quark-star to create flares before
colliding with the external shock.

If there are no, or only a few, slow shells from the quark-star that the 
black-hole jet collides with, the black-hole jet might not be 
slowed down significantly and it will catch up with
the external shock at an earlier stage. If this happens, a flare with not
too steep rise will result. However, after this flare, there will be no
flattening, and no bumps. The black-hole jet's collision with the 
external shock marks the end of the inner engine's contribution to the
observed afterglow light curve.
Thereafter, a ``classical'' afterglow decay is the result, with a possible jet
break due to non-isotropic ejecta creating the afterglow.

There is a correlation between the energy emitted in the flares and in
the bump (see \S~\ref{flaresbumpsection}), so that bursts 
with more significant flares will have less
pronounced bumps, and vice versa. This is assuming that a black-hole
was formed and an equal sized disk formed around the black-hole. A
small disk around the black-hole will lead to weak or no flaring and bumps.
In the case when no black-hole formed, no flares and no bump will be seen.

\subsection{Generic light curves in our model}

Here we discuss the eight different types of light curves that our model
can generate (Fig.~\ref{sixcases}).

{\bf Case 1.} This case has all the properties discussed
in the previous section $i.e.$ flares, flattening and a late time bump.
The flares are produced when a black-hole jet collides with slower shells
from the quark-star. The flattening is due to slower shells from the 
quark-star re-energizing the external shock, whereas the bump occurs
when the black-hole jet collides with the external shock.
This case is also shown in Fig.~\ref{flare-flattening}. 

{\bf Case 2.} This case shows a light curve with one or more flares and a flattening of
the X-ray afterglow light curve, but no bump. This will happen if the
later stages of the ejection from the quark-star produces slower shells.
These slower shells will re-energize the external shock, whereby the 
light curve is flattened. The accretion onto the black-hole launches 
an ultrarelativistic shell, and when this interacts with a slower shell
from the quark-star, an X-ray flare is seen. However, the merged shell 
from the black-hole and the quark-star jet is not fast enough to catch 
up with the external shock, so no bump is seen.

{\bf Case 3.} This case shows flattening and a bump, but no flares. This is because
the shell from the black-hole did not interact with any of the slower
shells from the quark-star. This indicates that the black-hole shell was
fairly slow. When the black-hole shell collides
with the external shock, a bump is seen. As before, slower shells in the
quark-star jet re-energizes the external shock, which flattens the
light curve.

{\bf Case 4.} This case has a bump, but no flare and no flattening. This occurs
when the quark-star jet does not produce any late time shells that
can flatten the light curve, and the shell from the black-hole can not
interact with any shells to create a flare. A bump is seen when the
shell from the black-hole jet collides with the external shock.

{\bf Case 5.} This case has a flare and a bump, but no flattening. This is rather
similar to case 4, however the quark-star jet emitted one or a few
low Lorentz factor shells that the black-hole jet interacted with. 
This happened before the late shells from the quark-star collided
with the external shock, so no flattening was created. GRB 050502B
is an example of a burst in this category.

{\bf Case 6.} This case shows flattening, but no flare and no bump. This occurs when
the quark-star emits low Lorentz factor shells that can flatten the
light curve, but no black-hole was formed, or the jet from the black 
hole was to weak too make any observable signatures.

{\bf Case 7.} This case is similar to case 6, but the quark-star did not emit late
time shells, so no flattening is seen.

{\bf Case 8.} This case shows flaring, but no flattening and no bump. This occurs when
the quark-star generates some slow shells and a black-hole is formed.
The jet from the black-hole interacts with the slow shells and produces
flaring. However, in this process the jet is slowed down enough that it
cannot make any significant impact on the external shock.

{\bf Case 1a.} In many observed afterglows the X-ray afterglow decays very
steeply just after the end of the gamma ray emission. We have explained this
as being due to the curvature effect, that is an off-axis gamma ray
emission, that in other directions is seen as the prompt emission. However,
in some bursts this steep decay is not observed. Instead the X-ray afterglow
declines gradually, with a power law similar to the decline at late times
($t>10^5$ seconds). We suggest that in this case the external shock sets in
earlier. There is still the possibility of having flares and flattening in
the same way as before, when the jet from the black-hole collides with
slower shells from the quark-star or the slower quark-star shells
re-energize the external shock.

\section{Light curve features: {\it Correlations}}
\label{flaresbumpsection}

\subsection{Anti-correlation between X-ray flares and bumps}

We have proposed that the jet launched by the black-hole is responsible
for the X-ray flares and bumps. The flares are produced when the black
hole jet collides with slower parts
of the jet from the quark-star. The bumps seen at later times in the
X-ray afterglow are due to the black-hole jet colliding with the 
external shock.

\begin{figure}
\includegraphics[scale=0.95]{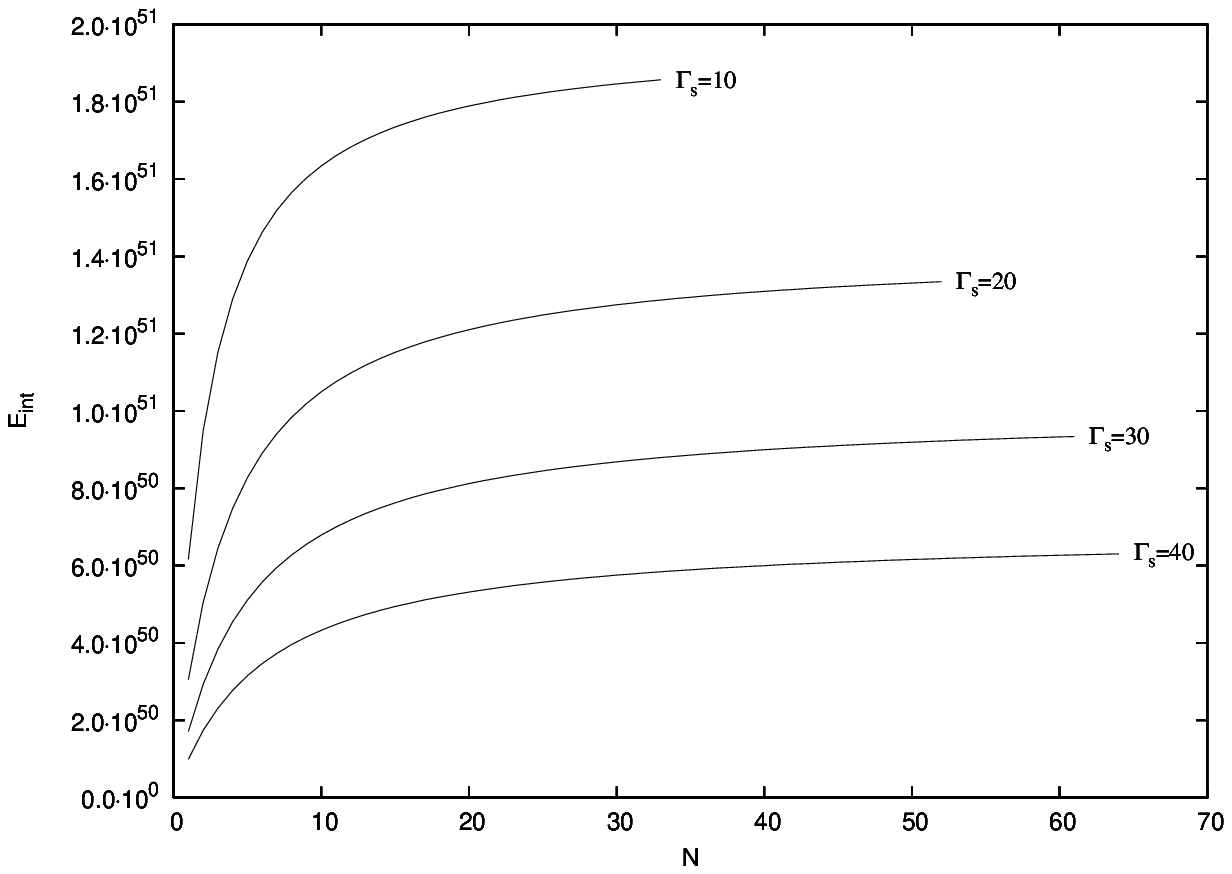}
\includegraphics[scale=0.95]{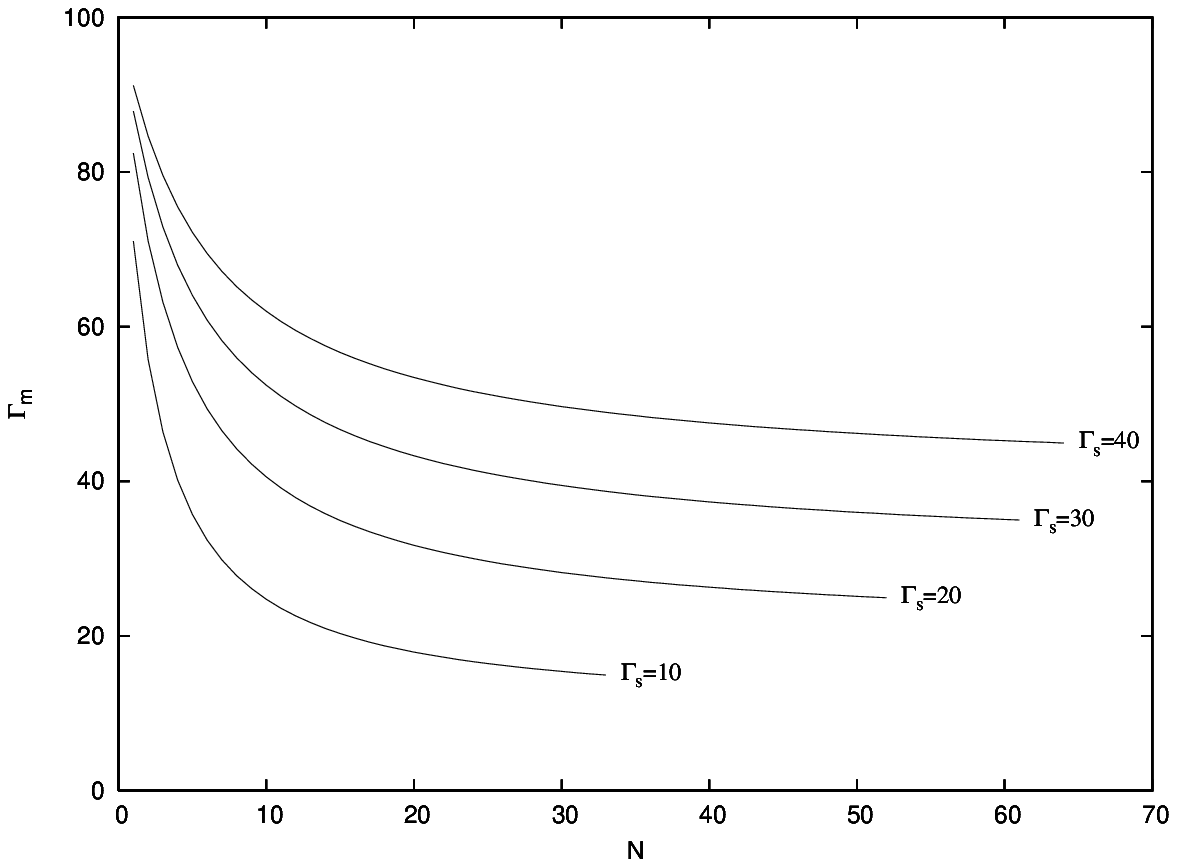}
\caption{Figure showing the total energy converted to internal energy 
versus number of
collisions (upper panel) and the Lorentz factor of the merged shell 
versus number of collisions (lower panel). In all cases the rapid 
shell has a Lorentz factor of 100 initially, and all the slow shells have
Lorentz factors of 10, 20, 30 or 40 respectively. The mass of the rapid shell is 
$3\times10^{28}$ g, and the mass of the slow shells are $3\times10^{27}$
g. The figures show that after 10 to 20 collisions a plateau
is reached and no more energy is converted. The Lorentz
factor reaches a plateau. In this plateau, collisions can still
occur, so the merged shell can gain mass. When the slower shells have
higher Lorentz factors, the merged shell ends up with a higher Lorentz
factor after the collisions, and the energy conversion is reduced. This
leaves more energy to be converted when this merged shell collides 
with the preceding quark-star external shock.}
\label{collisions}
\end{figure}

In Fig.~\ref{collisions} we explore the amount of energy converted 
and the resulting Lorentz factor of the merged shell when
the black-hole jet with Lorentz factor $\Gamma=100$ and mass 
$m=3\times10^{28}$ g collides successively with 
N quark-star shells. We show four different cases, when the quark-star
shells have Lorentz factors $\Gamma=10$, $20$, $30$, or $40$ and mass
$m=3\times10^{27}$ g (see Eqs~\ref{einteq} and \ref{gammameq} in the Appendix). 
In the first couple of collisions, the difference
in Lorentz factor is fairly large, so a lot of energy is converted.
However, after about 15 to 20 collisions, not much more energy is converted.
This is because the Lorentz factors of the colliding shells are not
very different. 
Several consecutive collisions will lead to several flares.
We also see that if the Lorentz factor
of the slow shells are high, the merged shell will get a higher Lorentz
factor but less energy will be converted.

The energy released in the bump depends on the kinetic energy of the
black-hole jet (or the merged jet that produced the flares) when it
collides with the external shock. If the jet is not very energetic,
i.e. a lot of energy was lost due to flaring, a weak bump will be 
seen. If not much energy was lost due to flaring,
because there were no slower shells from the quark-star, the bump
can be more pronounced.

To summarize: {\it Since both the flares and the bump are generated by
the black-hole jet, the maximum energy available to do so is the initial 
kinetic energy of the jet. If a large fraction of this energy is used to produce
flares, less is available for the bump, and vice versa.} 
This is in agreement with \citet{obrien06a,obrien06b}, who find that there
is a correlation between the energy output in flares and in what they
call the hump in observed GRB afterglows. The hump is the flattening
starting from about $t=10^3$ seconds and includes possible bumps.

\subsection{Correlation between prompt gamma energy and early X-ray afterglow
energy}

The total energy output from the quark-star inner engine is 
$\eta_{\rm QS}m_{\rm disk,QS}c^2\approx 2\times 10^{52}{\rm erg}$ for 
$m_{\rm disk,QS}\sim0.1M_\odot$. This energy is shared between the prompt
GRB\footnote{
For a $10\%$ radiative efficiency during shell collisions (as in the 
internal-external shocks model) in the quark-star phase this would 
imply that a maximum of about $2\times10^{51}$ erg is
released as synchrotron radiation.}, the external shock creating the decaying 
afterglow, and the re-energization of the external shock creating the
flattening of the early X-ray afterglow light curve. A GRB with low energy
in a prompt gamma-ray phase has more energy available for the early X-ray 
afterglow. Note however, that in the early X-ray afterglow more energy can
be emitted by the black hole jet interacting with the quark star ejecta,
as explained in the previous section.

\section{Case study}
\label{casestudysection}

In this section we will apply our model to a few observed X-ray afterglows
that illustrate the different properties of our model: flares, flattening
and bumps.

\subsection{GRB 050219A (case 2)}

The observed light curve with the X-ray telescope (XRT) and the 
Burst Alert telescope (BAT) do not look like they 
agree in this burst. This apparent discontinuity
was first reported in \citet{tagliaferri05} and
may be due
to an early X-ray flare \citep[as proposed by][]{obrien06a}
 after about 90 seconds. Since 
this flare is so early, it may either be a late 
internal shock produced purely by the quark-star, or possibly the
result of an early black-hole formation.
The early steep decay after this proposed flare is either the end of the 
flare or more likely due to the curvature effect \citep{meszaros06}. At later 
stages (from about 800 seconds) the light curve flattens and there are
some small flares. These are the late and slow parts of the quark-star jet that
re-energize the external shock. If we are right in our assumption that
there is an early flare after about 90 seconds, this would correspond
to case 2 in Fig.~\ref{sixcases}.

\subsection{GRB 050502B (case 5)}

GRB 050502B shows a very strong X-ray flare, starting after about 350 seconds 
and peaking around 700 seconds \citep{falcone06}. In addition, there 
is one (or possibly two) bumps seen at about 40000 to 50000 seconds after the GRB
trigger. The observed fluence of the flare is comparable or even bigger
than the GRB fluence. There are indications of substructures within the 
flare itself. 
The total energy of the flare (or the GRB) is not known. We find by 
using Eqs.~\ref{einteq} and \ref{gammameq} that a black-hole jet of 
$m_r=3\times 10^{28}$ g,
moving with Lorentz factor of $\Gamma_r=200$ and colliding with a
shell from the quark-star with Lorentz factor $\Gamma_s=10$ and mass 
$m_s=5\times 10^{27}$ gives an internal energy $E_{\rm int}=2.4\times 10^{51}$
ergs and a Lorentz factor of the merged shell of $\Gamma_m=96$. 
The mass of the slow shell from the quark-star may seem large, 
but this can be the result of collisions at earlier times. These
collisions may not have left observable flares if the difference in Lorentz
factors 
in the colliding shells were not large or these collisions
could have been part of the GRB itself. 
The black-hole must have ejected 
a fairly large amount of mass ($3\times10^{28}$ g), indicating that
the accretion disk must have been fairly large.

The bump(s) at $t>10^4$ seconds occur when the merged shell 
that created the flare catches up
with the external shock (as suggested in Falcone et al. 2006). The 
underlying decay of the afterglow is due to the external shock.
This burst corresponds to case 5 in Fig.~\ref{sixcases}.

\subsection{GRB 050421 (case 1, 2, 5 or 8)}

GRB 050421 shows two flares in the early X-ray afterglow, the first flare
peaking after 111 seconds and the second after 154 seconds \citep{godet06}. 
The duration of the
gamma ray emission was about 10 seconds. We remind the reader that these
flares are due to a jet from a black-hole colliding with slower parts of the
jet from the quark-star. The width of the first peak is about 10 seconds. 
This can be achieved if the quark-star emits a slow shell about
90 seconds following the first shell (c.f. eq.~\ref{pulsewidth}):
\begin{equation}
\delta t=\frac{L}{ac}=\frac{c\Delta t}{ac}=\frac{20}{2}=10~{\rm s}
\end{equation} 
($\Delta t$ being the time interval between the emission of the quark-star
shell and the black-hole shell). Alternatively, if the slower shell is
emitted at the end of this burst (after about 10 seconds) but the jet from
the black-hole has a significantly higher Lorentz factor, about 10 times
higher than the quark-star shell :
\begin{equation}
\delta t=\frac{L}{ac}=\frac{100 c}{10 c}=10~{\rm s}. 
\end{equation}
The second flare is created when the 
merged shell creating the first flare collides with another slow shell 
emitted from the quark-star. The later stages of this afterglow ($>10^3$
seconds) are not observed. Therefore we cannot say if there is any flattening
or bump. This corresponds to any one of cases 1, 2, 5 or 8 in 
Fig.~\ref{sixcases}.

Alternatively, these two flares could both be produced by internal shocks
from shells created in the late stages of the accretion onto the quark
star.

\subsection{GRB050401 (case 6 without steep decay)}

An example of a GRB without the early steep decline is GRB050401
\citep{depasquale06}.
The afterglows declines gradually with a
decay index $\alpha\approx0.65$ \citep{panaitescu06} until about 
4300 seconds. After this, the decay is steeper with a decay slope
$\alpha\approx 1.39$. We ascribe the behavior until about 4300 seconds
as flattening due to slower quark star shells re-energizing the external 
shock. There is no clear evidence of any flaring in this afterglow, in 
which case no black-hole jet interacted with the quark star jet or the
external shock. This corresponds to case 6 without the initial steep 
decay.

The isotropic equivalent energy of this burst is about 
$E_{\gamma, iso}\sim3\times10^{53}$ erg. Assuming a beaming angle of 
$5^\circ$, this gives a total energy in gamma rays of about 
$4\times10^{50}$ erg (for a bipolar jet). \citet{zhang06} found 
that almost five times
as much energy was injected in this burst during the afterglow than
during the prompt gamma radiation, meaning about $2\times 10^{51}$ ergs was
released in the early X-ray afterglow. In our model, this corresponds to
$0.1 M_\odot$ disk being accreted onto a quark star with $10\%$ of the accreted
energy ejected into a bipolar jet, and only about $2\%$ of the jet energy goes
into radiation. Most of the jet energy (the remaining $98\%$)
is used to re-energize the external shock. 

\section{Discussion}
\label{discussionsection}

\subsection{Hypernova-GRB association in our model}

The formation of the quark-star through a quark nova (between stage1 and 2)
shortly after the core collapse
or directly in the core collapse releases an extra amount of energy that can
re-energize the SN ejecta, reminiscent of a hypernova 
\citep{ouyed02, keranen05}.
GRBs seem to be associated with hypernovae 
\citep[i.e. GRB 030329;][]{hjorth03}, \citep[GRB 980425;][]{galama98}, however, the opposite is not
always true \citep[i.e. SN 2002ap;][although a lack of GRB
observation does not necessarily mean that no GRB occurred]{mazzali02}. 
In order to get a hypernova, the energy from the conversion to quark matter
is necessary. In our model, the GRB is produced by the jet launched by
accretion onto the quark star. If a black hole is formed directly in the
collapse (no quark star stage), there will be no hypernova. Accretion onto
the black-hole can launch a jet, and internal shocks in the jet can produce
a GRB (see \S~\ref{noqsstagesubsection}). However, since there is only one stage,
there can be no late activity.

\subsection{SN-less GRBs in our model}

GRB060614 and GRB060505 were apparent nearby, long duration GRBs that showed
no sign of a supernova, leading to the term SN-less GRBs \citep{fynbo06}.
In the litterature possible explanation for this have been
that not enough $^{56}Ni$ is formed in the explosion \citep{tominaga07}, or 
that this is a very long and energetic ``short'' burst \citep{zhang07}. Here 
we give a possible interpretation of this phenomena within our model.
We suggested above that the energy released by the quark nova (formation of
a quark-star) could potentially lead to a hypernova. However, if the star
is too massive to explode even with the extra energy released from the
quark nova it will instead collapse and no supernova/hypernova will 
be seen. 
We assume that a hyperaccretion disk can form around the quark star while
most of the star is trying to explode\footnote{This scenario is initially 
similar to that leading to a GRB with an associated hypernova, only 
that in this case the ejected material does not have enough energy to 
escape the star.},
and as before accretion onto the quark-star powers an ultrarelativistic 
jet giving rise to a GRB. In this scenario a black-hole is the most 
likely final outcome: either the envelope falls back directly onto
the compact core, or it feeds to the accretion disk.
In the first case a black hole is formed but no ultrarelativistic jet is 
launched from the accretion onto it and hence no flaring or bump activity
is seen in the early X-ray afterglow. In the second case the quark-star
collapses to a black hole when it has accreted enough mass and continued
accretion onto the black hole will launch an ultrarelativistic jet that
can give rise to flaring and bump activity seen in the early X-ray afterglow.

\subsubsection{Black-hole formation without a quark-star stage}
\label{noqsstagesubsection}
In the event that a black-hole is formed immediately following the
core collapse, a collapsar scenario can occur if there is sufficient angular
momentum present. Simulations of this scenario by De Villiers et al. (2005)
show that an ultrarelativistic jet is launched that can give rise to
internal shocks and a GRB. In addition to the jet, a coronal wind was 
launched that could explode the remainder of the star. However, the energy 
of this wind was of the order $10^{48}-10^{49}$ erg, which is not enough to 
explain a supernova. Hence this could be another possibility for explaining 
supernova-less GRBs, as discussed above. The duration 
of the accretion process is short (accretion rate of the order $1M_\odot~{\rm s^{-1}}$ 
leading to a rather short duration GRB). 

An alternative occurs when the coronal wind is not strong enough to explode
the star. The star will then collapse, and assuming that it has high angular
momentum, it will form a massive accretion disk around the black-hole. The
accretion rate onto the black-hole is probably of the same order as with a
smaller disk, however with much more mass to accrete this could lead to a
somewhat longer duration jet and GRB. This resembles the original failed
supernova discussed by Woosley (1993).

Both cases discussed above lead to a one stage model resulting in the
absence of late time activity. Therefore, there will be no flares or bumps 
in the afterglow light curve, but flattening is still possible.

\subsection{Collapsar vs. accreting quark-star as inner engine for GRBs}

In this section we will briefly discuss the differences and similarities 
between our model and the collapsar 
model \citep{woosley93} for the inner engine for GRBs.

{\it Similarities:} Both the collapsar and our model assume the death 
of a massive, rapidly rotating star as the triggering mechanism. Both
models launch an ultrarelativistic jet, in which internal shocks 
accelerate electrons that produce synchrotron radiation. As this jet
is decelerated by the ISM, an external shock forms in which electrons 
again are accelerated and emit synchrotron emission that is responsible
for the afterglow.

{\it Differences:} The collapsar model assumes that the compact core left
behind in the collapse of the core of the progenitor star quickly collapses to a
black-hole, whereas in our model this compact core collapses to a quark
star instead. This opens up some interesting possibilities. The accretion
rate into a black-hole and a quark-star are totally different. 
Accretion into a black-hole is a very rapid process, with accretion
rates of the order of one solar mass per second. Accretion onto a quark 
star on the other hand is much slower (about $10^4$ times slower). This can 
therefore easier explain the long duration ($\sim 1000$ seconds) seen 
in some bursts. A main focus of this paper is the flaring often seen in 
the X-ray afterglow, which is indicative of the engine being restarted. 
With a quark-star the possibility of it collapsing to a black-hole exists,
which explains how the inner engine can be restarted leading to the observed
features in the early X-ray afterglow. 

\section{Summary and conclusion}
\label{summarysection}
To summarize, the properties of the three stages of the inner 
engine in our model are:
\begin{itemize}
\item The process is initiated by the collapse of the iron core of 
a massive, rapidly rotating star.
\item The collapsed core will either leave a neutron star or a quark-star
behind.
\item If a neutron star is left behind, it can collapse to a quark-star
at a later stage, creating a delay between the supernova and the GRB.
\item Fallback material from the supernova and the quark nova forms a disk 
around the quark-star. Accretion onto the quark-star generates the 
GRB by powering an ultrarelativistic jet. Internal shocks in this jet 
create the GRB.
\item Accretion continues, but at some point it cannot heat up the star 
sufficiently. This halts the emission of shells, ending the GRB.
\item When the quark-star has accreted too much material, it collapses
into a black-hole. Further accretion onto the black-hole launches an
ultrarelativistic jet. 
\end{itemize}
The emission features in our model can be summarized as follows:
\begin{itemize}
\item Early, steep decay in the X-ray afterglow is due to the curvature
effect
(this is not specific to our model, but rather a generic feature in many models).
\item Flares are created by interaction between the jet from the 
accretion onto a black-hole and slower parts from the jet from the quark-star.
\item Re-energization of the external shock (seen as flattening of the
X-ray afterglow light curve) is due only to the jet from the quark-star. 
Slower parts of this jet re-energizes the external shock.
\item When the jet from the black-hole collides with
the external shock from the GRB, a bump is seen in the afterglow light curve.
\end{itemize}

In conclusion, we have presented a three stage (effectively two stage) 
model for the inner engine
for GRBs involving a neutron star phase, followed by a quark star phase then by
a black hole phase. This model seems to account for the observed prompt gamma ray
emission, as well as the features of the early X-ray afterglow and as such
warrants further study.

\acknowledgements
We would like to thank J. Hjorth, M. Lyutikov M. Cummings, W. Dobler, 
D. Leahy and B. Niebergal as well as the anonymous referee for helpful 
remarks.

\appendix

\section{Internal-external shocks model}

\citet{piran05} 
shows that the pulse width is proportional to the separation between 
two shells:
\begin{equation}
\delta t\approx R_s/2a\Gamma^2c=L/ac {\rm \hspace{1cm} equal~mass~shells},
\label{pulsewidth}
\end{equation}
where $R_s$ is the distance at which the collision takes place, $a$ is the
ratio between the fast and slow shells Lorentz factor, $\Gamma$ the 
Lorentz factor of the slow shell, L the separation of the shells and c
the speed of light.
Assuming
that the colliding shells have equal mass, and the faster has a Lorentz
factor two times the slower shell, the separation between the two shells
must be $3\times10^{13}$ cm in order to have a pulse width of 500 seconds.
This corresponds to a thousand seconds separation between the emission of the
two shells.

The shells will collide at a distance 
\begin{equation}
R_s\approx 2\Gamma^2L
\end{equation}
where $\Gamma$ refers to the Lorentz factor of the slower shell. For 
$\Gamma=20$ and a separation of $3\times10^{13}$ cm as before, the 
collisions occur at $R_s=2\times10^{16}$ cm. If the Lorentz factors
are too big, the external shocks will occur before internal shocks could
occur. 
If the photon energy is large enough to produce $e^+e^-$ pairs, the
Lorentz factors of the shells have to be above a hundred to overcome
the compactness problem. However, if the shocks creates X-rays, the
photons cannot create $e^+e^-$ pairs, and there is no lower bound on the
Lorentz factor.

The internal energy of the merged shell is given by:
\begin{equation}
E_{\rm int}=m_rc^2(\Gamma_r-\Gamma_m)+m_sc^2(\Gamma_s-\Gamma_m),
\label{einteq}
\end{equation}
where $m_r$ is the mass of the rapid shell, $m_s$ of the slow shell, 
$\Gamma_r$ is the Lorentz factor of the rapid shell, $\Gamma_s$ is 
the Lorentz factor of the slower shell and $\Gamma_m$ of the merged shell:
\begin{equation}
\Gamma_m=\sqrt{\frac{m_r\Gamma_r+m_s\Gamma_s}{m_r/\Gamma_r+m_s/\Gamma_s}}.
\label{gammameq}
\end{equation}
A certain fraction of this energy will be emitted as radiation. In 
order to have an efficient collision (a collision in which a lot of 
energy is converted to internal energy), the masses must be similar, and the
Lorentz factor of the fast shell must be at least twice that of the slower
shell.
For a mass of the rapid shell of about $3\times10^{28}$ g, which is not
unlikely in a black-hole jet \citep[$\dot{M}\sim10^{-5} M_{\rm disk}/s$ in][]{devilliers05}, and a mass of the 
slow shell of $3\times10^{27}$ g, Lorentz factor of the rapid shell of 
$\Gamma_r=65$, and Lorentz factor of the slow shell of 
$\Gamma_s=20$, which gives $\Gamma_m=57$ and an internal energy of about
$10^{50}$ erg in a collision.

\clearpage

\begin{thebibliography}{}

\bibitem[Alcock et al.(1986)]{alcock86} Alcock, C., Farhi, E., \& Olinto, A. 1986, ApJ, 310, 261

\bibitem[Bagchi et al.(2006)]{bagchi06} Bagchi, M., Ouyed, R., \& Staff, J. E., Ray, S., Dey M., Dey J., 2006, astro-ph/0607509.

\bibitem[Beloborodov et al. (2000)]{beloborodov00} Beloborodov, A. M., Stern, B. E., \& Svensson, R., 2000, ApJ, 535, 158.

\bibitem[Burrows et al. (2005)]{burrows05}Burrows, D. N., et al., 2005, Science, 309, 1833

\bibitem[Dai et al.(2006)]{dai06} Dai, Z. G., Wang, X. Y., Wu, X. F., Zhang, B., 2006, Science, 311, 1127

\bibitem[De Villiers et al.(2005)]{devilliers05} De Villiers, J. P., Staff, J. E., \& Ouyed, R., 2005, astro-ph/0502225.

\bibitem[Della Valle(2006)]{dellavalle06}Della Valle, M. ``16th Annual October Astrophysics Conference in Maryland'', eds. S. Holt, N. Gehrels and J. Nousek, AIP Conf. Procs. astro-ph/0604110

\bibitem[De Pasquale et al.(2006)]{depasquale06} De Pasquale, M., et al., 2006, MNRAS, 365, 1031

\bibitem[Falcone et al.(2006)]{falcone06} Falcone, A. D., et al., 2006, ApJ, 641, 1010.

\bibitem[Frank et al.(1992)]{frank92} Frank, J., King, R., Raine, \& D. J. 1992, Accretion Power in Astrophysics (Cambridge: Cambridge Univ. Press)

\bibitem[Fryer et al.(1996)]{fryer96} Fryer, Ch. L., Benz, W., Herant, M. 1996, ApJ, 460, 801

\bibitem[Fynbo et al.(2006)]{fynbo06} Fynbo, J., et al. 2006, Nature, 444, 1047

\bibitem[Galama et al.(1998)]{galama98} Galama, T. J., et al. 1998, Nature, 395, 670

\bibitem[Gehrels et al.(2004)]{gehrels04} Gehrels, N., et al. 2004, ApJ, 611, 1005

\bibitem[Godet et al.(2006)]{godet06} Godet, O., et al., 2006, A\&A, 452, 819

\bibitem[Harko \& Cheng(2002)]{harko02} Harko, T., Cheng, K. S., 2002, A\&A, 385, 947

\bibitem[Hjorth et al.(2003)]{hjorth03} Hjorth, J. et al., 2003, Nature, 423, 847

\bibitem[Itoh(1970)]{itoh70} Itoh, N, 1970, PThPh, 44, 291

\bibitem[Iwazaki (2005)]{iwazaki05} Iwazaki, A. 2005, PhRvD, 72, 114003

\bibitem[Ker{\"a}nen et al.(2005)]{keranen05} Ker{\"a}nen, P., Ouyed, R., \& Jaikumar, P., 2005, ApJ, 618, 485

\bibitem[Kluzniak \& Ruderman(1998)]{kluzniak98} Klu{\'z}niak, W., Ruderman, M., 1998, ApJL, 505, 113

\bibitem[Kobayashi et al.(1997)]{kobayashi97} Kobayashi, S., Piran, T., \& Sari, R., 1997, ApJ, 490, 92

\bibitem[Kumar \& Panaitescu(2000)]{kumarpanaitescu00} Kumar, P. \& Panaitescu, A. 2000, ApJ, 541,L51

\bibitem[Liang et al.(2006)]{liang06} Liang, E. W., et al., 2006, ApJ, 646, 351

\bibitem[Mazzali et al.(2002)]{mazzali02} Mazzali, P., et al., 2002, ApJ, 572, L61

\bibitem[M{\'e}sz{\'a}ros(2006)]{meszaros06} M{\'e}sz{\'a}ros, P. , 2006, ``16th Annual October Astrophysics Conference in Maryland'', eds. S. Holt, N. Gehrels and J. Nousek, AIP Conf. Procs. astro-ph/0601661

\bibitem[Narayan et al.(1992)]{narayan92}Narayan, R., Paczynski, B., \& Piran, T., 1992, ApJ, 392, L83

\bibitem[O'Brien et al.(2006a)]{obrien06a} O'Brien, et al., 2006a, ApJ, 647, 1213

\bibitem[O'Brien et al.(2006b)]{obrien06b} O'Brien, P. T., Willingale, R., Osborne, J. P., \& Goad, M. R., 2006b, NJPh, 8, 121.

\bibitem[Ouyed et al.(2002)]{ouyed02} Ouyed, R., Dey, J., \& Dey, M. 2002, A\&A, 390, 39

\bibitem[Ouyed et al.(2005)]{ouyed05} Ouyed, R., Rapp, R., \& Vogt, C., 2005, ApJ, 632, 1001 (ORV).

\bibitem[Ouyed et al.(2006)]{ouyed06} Ouyed, R., Niebergal, B., Dobler, W., \& Leahy, D., 2006, ApJ, 653, 558.

\bibitem[Panaitescu et al.(2006)]{panaitescu06} Panaitescu, A., M{\'e}sz{\'a}ros, P., Burrows, D., Nousek, J., Gehrels, N., O'Brien, P., \& Willingale, R. 2006, MNRAS, 369, 2059

\bibitem[Piran(1999)]{piran99}Piran, T., 1999, PhR, 314, 575

\bibitem[Piran(2005)]{piran05} Piran, T., 2005, Rev. Mod. Phys., 76, 1143.

\bibitem[Popham et al.(1999)]{popham99} Popham, R., Woosley, S.E., \& Fryer, Ch. 1999, ApJ, 518, 356.

\bibitem[Rhoads(1999)]{rhoads99}Rhoads, J. E. 1999, ApJ, 525, 737.

\bibitem[Ruderman et al.(2000)]{ruderman00} Ruderman, M. A., Tao, L., Klu{\'z}niak, W., 2000, ApJ, 542, 243

\bibitem[Sari \& M{\'e}sz{\'a}ros(2000)]{sarimeszaros00} Sari, R. \& M{\'e}sz{\'a}ros, P. 2000, ApJ, 535, L33.

\bibitem[Staff et al.(2006)]{staff06} Staff, J., Ouyed, R., \& Jaikumar, P. 2006, ApJ, 645, L145 

\bibitem[Tagliaferri et al.(2005)]{tagliaferri05} Tagliaferri, G. et al., 2005, Nature, 436, 985

\bibitem[Tominaga et al.(2007)]{tominaga07} Tominaga, N., Maeda, K., Umeda, H., Nomoto, K., Tanaka, M., Iwamoto, N., Suzuki, T. \& Mazzali, P. A., 2007, ApJ, 657, L77.

\bibitem[Usov(1992)]{usov92} Usov, V. V. 1992, Nature, 357, 472

\bibitem[Vogt et al.(2004)]{vogt04} Vogt, C., Rapp, R., \& Ouyed, R. 2004, Nuc. Phys. A, 735, 543.

\bibitem[Wheeler et al.(2000)]{wheeler00} Wheeler, J. C., Yi, I., H{\"o}flich, P., Wang, L., 2000, ApJ, 537, 810.

\bibitem[Woosley(1993)]{woosley93} Woosley, S.E. 1993, ApJ, 405, 273.

\bibitem[Zhang et al.(2006)]{zhang06} Zhang, B., Fan, Y.Z., Dyks, J. Kobayashi, S., M{\'e}sz{\'a}ros, P., Burrows, D.N., Nousek, J.A., \& Gehrels, N. 2006, ApJ, 642, 354.

\bibitem[Zhang et al.(2007)]{zhang07} Zhang, B, Zhang, B. B., Liang, E. W., Gehrels, N., Burrows, D. N. \& M{\'e}sz{\'a}ros, P. 2007, ApJ, 655, L25.

\bibitem[Zou et al.(2006)]{zou06} Zou, Y. C., Dai, Z. G., \& Xu, D. 2006, ApJ, 646, 1098.

\end{thebibliography}
\end{document}